\documentstyle[12pt]{article}
\newcommand{\be}{\begin{equation}}
\newcommand{\bea}{\begin{eqnarray}}
\newcommand{\ee}{\end{equation}}
\newcommand{\eea}{\end{eqnarray}}
\newcommand{\dsa}{{d \sigma^{\Lambda (s)} (s')  \over dx d \Omega}(x,Q^2)}
\newcommand{\dsb}{{d \hat \sigma^{ a(\lambda)} (s')  \over dy d \Omega}}
\newcommand{\dsc}{{d \sigma^{\Lambda(\downarrow)}   \over dx d \Omega}
-{d \sigma^{\Lambda(\uparrow)}  \over dx d \Omega}}
\newcommand{\dsd}{{d \hat \sigma^{a(\downarrow)} \over dy d \Omega}
-{d \hat \sigma^{ a(\uparrow)}  \over dy d \Omega}}

\newcommand{\dsg}{{d \sigma^{\Lambda(\downarrow)}(\downarrow) 
\over dx d \Omega}
\!-\!{d  \sigma^{\Lambda(\uparrow)}(\downarrow)  \over dx d \Omega}}
\newcommand{\dsf}{{d \hat \sigma^{a(\downarrow)}(\downarrow) 
\over dy d \Omega}
\!-\!{d \hat \sigma^{a(\uparrow)}(\downarrow)  \over dy d \Omega}}
\begin{document}
\begin{titlepage}
\vspace*{\fill}
\begin{center}
{\Large \bf The Polarised $\Lambda$ production in QCD}\\[1cm]
{\bf V. Ravindran}\\
{\em Theory Group, Physical Research Laboratory, Navrangpura \\
Ahmedabad 380 009, India}\\
\end{center}
\vspace{2cm}
\begin{abstract}
The $Q^2$ evolution of polarised parton fragmentation functions is
discussed using Altarelli-Parisi evolution equations.  The first moments
of both polarised quark and gluon fragmentation functions are shown to
behave in a similar fashion at very high energies.  
This analysis is applicable to any hard processes involving
production of polarised hadrons.  The polarised
$\Lambda$ hyperon production in $e^+~e^-$ annihilation where
this can be realised is considered.  We present a complete 
$\alpha_s(Q^2)$ corrections to the asymmetries discussed in the paper of
Burkardt and Jaffe which demonstrates the extraction of various polarised 
fragmentation functions.  To this order, these corrections 
are found to be scheme dependent similar to that of structure functions.  
\end{abstract}
\vspace*{\fill}
\end{titlepage}

\section{\bf Introduction}
The theory of strong interaction physics is well understood
within perturbative Quantum Chromodynamics(PQCD) $\cite {MUTA}$.  
The PQCD describes the dynamics of quarks and gluons at very high energies.
The laboratory experiments involving
hadrons either in the initial state or final state can be understood
by treating these quarks and gluons as asymptotic
states and by safely using the perturbative techniques thanks to
asymptotic freedom.  In other words,
the hadronic cross sections can be well expressed in terms of partonic
cross sections.  The Parton Model (PM) \cite {ALT} has been a successful 
model to tie up these perturbatively
calculable parton level cross sections and the experimentally measured
cross sections.  This involves the introduction of certain probability
distributions which can not be calculable 
in PQCD but are just inputs of the model.
Though they are not calculable, they are found to be universal in the 
sense that
they are process independent and the evolution in terms of the scale is well 
known.  In other words, if these probability distributions are measured
in an experiment at a scale, say $Q^2$ which is above the 
$\Lambda_{QCD}$ where
the QCD perturbation theory is reliable, then these distributions can
be used as inputs to predict the rates or cross sections of different
experiments which involve these distributions.  
In this sense they are something to do with the hadrons 
participating in the process and are process independent.
Also, using the well known Altarelli-Parisi evolution equation satisfied
by these distributions, one can find out how these distributions change as 
the scale changes and thereby use them appropriately for the experiments done
at various energies.  

There are two types of such probability distributions one comes across
in hadron physics.  The well understood among them is the parton probability
distribution.  It is generally denoted by $f_{a(h) /H(s)}(x,Q^2)$ and the
meaning of it is the probability of finding a parton of type $a$ with
polarisation $h$ inside the hadron $H$ of polarisation $s$ with momentum
fraction $x$ of the hadron and the scale $Q^2$.  These distributions
are usually measured in Deep Inelastic Scattering(DIS) experiments.
The universality of these distributions and the predicted evolution 
in terms of the scale $Q^2$ are proved experimentally $\cite {ALT}$.  The other
distribution is nothing but the fragmentation function.
The fragmentation functions are mirror image of the parton probability
distributions.  These functions are generally denoted by
$D_{a(h)}^{H(s)}(x,Q^2)$.  They measure
the probability that a parton of type $a$ with polarisation $h$ fragments
into a hadron of type $H$ with polarisation $s$ and carries away
$x$ fraction of the parton's momentum at a scale $Q^2$ \cite{BF}.  
These functions are
usually measured in $e^-$ $e^+$ $ \rightarrow H~X$ experiments.
The universality and the $Q^2$ evolution of unpolarised functions
(which does not contain any information about the spin of the hadrons
produced) are well understood.  

In recent years, there has been several interesting works to understand
these fragmentation functions. 
These include the measurement of the fragmentation functions for 
charged and neutral pions and kaons $\cite {MAT}$.
The QCD inspired Parton Model analysis sublamented with
Altarelli-Parisi evolution equations for these functions
is shown to explain the pion and kaon production rates at leading and 
next-to-leading order $\cite {BKK}$.   The another field of current
interest is the inclusive $\Lambda$ production at $e^-$ $e^+$ scattering
experiments $\cite {LUN}$.  This is important as 
$\Lambda~$s are produced copiously
and also it is easy to detect them.  On the theoretical side,
the study of polarisation effects in the production cross sections 
is more interesting.  As we know, at low energies the unpolarised
$e^-$ $e^+$ scattering will not produce longitudinally 
polarised $\Lambda~$s (we do not attempt to review the 
transversely polarised $\Lambda~$s produced 
in unpolarised scattering experiments) but
when the center of mass energy crosses the $Z$ threshold, then
due to parity violating effects, the produced $\Lambda~$s can
naturally be polarised.  Recall that these are produced due to
the fragmentation of polarised partons produced in the scattering.
Unfortunately, it is not clear how these partons fragment into
these hadrons.  There exists only model calculations which can
tell us how much of the parton spin is transferred to $\Lambda$.
In the naive quark model, the only contribution to polarised
$\Lambda$ is due to strange quark and the contribution
coming from other partons is identically zero.  As we know
this naive picture breaks down due to complicated
QCD effects of both perturbative and non-perturbative origins.
Along this line of thought there has been an interesting work
by Burkardt and Jaffe $\cite {BUR}$ who have charted 
out an experimental programme
to measure polarised $u$ and $d$ quark fragmentation functions 
in addition to $s$ quark fragmentation function.  The non-zero 
value for $u$ and $d$ quark
fragmentation functions will invalidate the quark model picture.
More recently the references $\cite {RAV1},\cite {RAV2}$ discuss the
importance of gluons in the polarised $\Lambda$ production.
The analysis in the reference $\cite{RAV1}$ is purely 
based on the AP evolution equations satisfied by
the polarised fragmentation functions of quarks and gluons.
It has been shown that the gluons play significant role
when the scale at which the experiment is done is very high.
In the massive gluon scheme, it has been demonstrated \cite {RAV2}
that the gluons contribute to polarised lambda production.
Since the fragmentation functions are defined in this scheme,
the gluonic contribution to polarised $\Lambda$ is
scheme dependent.  The situation is very similar to
the gluonic contribution to polarised structure function
measured in DIS.  This paper extends the previous analysis \cite {RAV2}
to include the $Z$ boson exchange in order to completely extract
various parton fragmentation functions.  

The paper is organised as follows.  In the second chapter we discuss the
importance of polarised gluons in the $\Lambda$ production
using AP evolution equations.  We then systematically compute 
the QCD corrections to the asymmetries defined below
which are useful to extract various partonic contributions to
the $\Lambda$ production.  The results are presented in the chapter 3.
We finally conclude in the chapter 4. 
The appendix contains the relevant details of the 
results presented in the chapter 3.  

\section{\bf The $Q^2$ Evolution of the Polarised Fragmentation Functions}
In the parton model, the production cross section for $\Lambda$
with polarisation $s$ in $e^-$ $e^+$ scattering is related to 
that of a parton and the probability 
that the parton fragments into $\Lambda$.   In other words,
\be
\dsa = \sum_{a,\lambda} \int_x^1 {dy \over y} \dsb D_{a(\lambda)}^{\Lambda(s)}
       (x/y,Q^2)
\ee
The left hand side is 
the hadronic cross section for the production
of polarised $\Lambda$ within a solid angle $\Omega$ with respect to the 
beam direction in $e^-$ $e^+$ scattering.  Here $s$, $s'$ and $\lambda$ 
are the helicities of $\Lambda$, electron and parton respectively.
The Bjorken like variables $x=2 p_{\Lambda}.q/Q^2$ and $y=2 p_a.q/Q^2$ are
scaling variables in the hadron and parton levels respectively
with $q^2=Q^2$, 
$p_{\Lambda}$,$p_a$ and $q$ being the momenta of produced hadron, 
parton and the intermediate vector boson($\gamma$ or $Z$) respectively.  
The production cross section for a parton with helicity
$\lambda$ (right hand side of the equation)
is completely calculable in PQCD.  On the other hand, the 
probability that a parton of helicity $\lambda$ fragments into
$\Lambda$ of helicity $s$, $D_{a(\lambda)}^{\Lambda(s)}(x/y,Q^2)$, 
is not calculable due to non-perturbative
mechanism involved in the fragmentation region.  Hence, these
fragmentation functions are just inputs of the model
which can be extracted from the experiment.  Though they are
not calculable within the realm of PQCD, they are universal in the
sense that they are process independent.   More interestingly
their evolution in terms of $Q^2$ and $x$ are completely
governed by the well known Altarelli-Parisi evolution equations.

In the polarised experiments, the hadronic cross section is directly
proportional to polarised parton fragmentation functions defined as
\be
\Delta D_a^\Lambda(x,Q^2)= D_{a(\uparrow)}^{\Lambda(\uparrow)}(x,Q^2)- 
D_{a(\uparrow)}^{\Lambda(\downarrow)}(x,Q^2)
\nonumber
\ee

Let us first analyse the $Q^2$ evolution of these fragmentation 
functions using Altarelli-Parisi evolution equations.  
The evolution equations are well known and are given by
\be
{d \over dt} \Delta D_{q_i}^\Lambda(x,t) = {\alpha_s(t) \over 2 \pi}
\int_x^1 {dy \over y} \left[ \Delta D_{q_i}^\Lambda(y,t) \Delta P_{qq}(x/y)
+ \Delta D_g^\Lambda(y,t) \Delta P_{gq}(x/y)\right]
\label{apeqnq}
\ee
\be
{d \over dt} \Delta D_g^\Lambda(x,t)\! =\! {\alpha_s(t) \over 2 \pi}
\!\int_x^1\!\! {dy \over y} \left[ \sum_{j=1}^{2f}\!\Delta D_{q_i}^\Lambda(y,t) 
\Delta P_{qg}(x/y)\!\! +\!\! \Delta D_g^\Lambda(y,t) \Delta P_{gg}(x/y)\right]
\label{apeqn}
\ee
where $t=\log(Q^2/\Lambda^2)$ and $\alpha_s(t)$ is the strong coupling
constant.  Here, $\alpha_s(t) \Delta P_{ab}(y) dt$ 
is the probability density of finding
a parton of type $a$ at the scale $t+dt$ with momentum fraction
$y$ inside the parton of type $b$ at a scale $t$.  
The splitting functions are given by
\bea
\Delta P_{qq}(z)&=&C_2(R)\left({1+z^2 \over (1-z)_+}+ {3 \over 2} \delta(1-z) 
\right) 
\nonumber\\
\Delta P_{gq}(z)&=&C_2(R) \left( {1- (1-z)^2 \over z}\right)
\nonumber\\
\Delta P_{qg}(z)&=&{1\over 2} (z^2-(1-z)^2)
\nonumber \\
\Delta P_{gg}(z)&=&C_2(G)\left( (1+z^4) \left({1\over z} + {1\over (1-z)_+}
\right) \right. \nonumber \\
&&\left. -{(1-z)^3 \over z} + \left( {11 \over 6} - {2 \over 3} 
{T(R) \over C_2(G)}\right) \delta(1-z) \right)
\nonumber
\eea
Here, $C_2(R)= (N^2-1)/2N$, $C_2(G)=N$ and $T(R)=f/2$ with $N=3$ for 
$SU(3)$ and $f$ being the number of flavours $\cite {MUTA}$.  In 
the above equations, usual $+$ prescription has been used 
to regulate $z \rightarrow 1$ singularity.  
Notice that the above equations are similar to the AP equations
satisfied by the polarised parton distribution functions but for
the interchange in $\Delta p_{qg}$ and $\Delta p_{gq}$.
The reason for this is as follows:
The emission of a quark(gluon) from a quark
(gluon) only affects the probability of quark (gluon) fragmenting 
into hadron.  Hence the splitting functions
$\Delta P_{qq}$ and $\Delta P_{gg}$ are unaffected.
On the other hand, the emission of a quark from a gluon 
changes the probability of the gluon fragmenting into hadron.  
Similarly, the emission of a gluon from a quark affects the probability
of quark fragmenting into hadron.  That is why, 
the splitting functions $\Delta P_{qg}$ and $\Delta P_{gq}$ are interchanged.
As these equations can easily be solved in
the Mellin space,  we define
\bea
\Delta D_a^\Lambda(n,t)&=&\int_0^1 x^{n-1} \Delta D_a^\Lambda(x,t) dx
\nonumber \\
\Delta P_{ab}(n)&=&\int_0^1 x^{n-1} \Delta P_{ab}(x) dx
\eea
The complete solution for the $n$th moment is not illuminating.
Recall that the first moments of the polarised parton distributions
are interesting as they are related to the spin content of the
polarised hadron and the measurement of them will tell us
how the hadron spin is shared among the partons.
In the same spirit, we here look at the first moment of polarised
parton fragmentation functions.  
The measurement of it for various hadrons will tell us how the 
parton helicity is distributed among the produced hadrons.
But it is experimentally a hard task.
From eqn.(\ref {apeqn}), we find that
the first moment of the polarised gluon fragmentation function to 
order $\alpha_s(t)$ satisfies a simple first order differential 
equation, that is
\be
{ d \over dt} \Delta D_g^\Lambda(t) = \alpha_s(t) \beta_0 \Delta D_g^\Lambda(t) 
\label{apg}
\ee
where $\beta_0=(11 C_2(G)- 4 T(R))/12 \pi$.
The solution to the above equation can be found very easily 
using renormalisation group(RG) equation for the QCD coupling constant,  
\be
{d \over dt}\alpha_s(t)= - \beta_0 \alpha_s^2(t)
\label{alrg}
\ee
From eqns.(\ref {apg}, \ref {alrg}),
we obtain an interesting behaviour of first moment of gluon 
fragmentation function:  the product of the first moment of
polarised gluon fragmentation function times the strong coupling constant
is scale independent to order $\alpha_s^2(t)$,
\be
{d \over dt} (\alpha_s(t) \Delta D_g^\Lambda(t))= 0 (\alpha_s(t)^3)
\label{aldg}
\ee
In other words, to order $\alpha_s^2(t)$, $\Delta D_g^\Lambda$ increases 
as the scale $t$ increases, i.e
\be
\Delta D_g^\Lambda(t) = K \log \left({Q^2 \over \Lambda^2}\right)
\ee
where $K$ is some constant.  It is worth recalling that the 
counter part of such a relation for polarised gluon 
distribution function exists and has opened up a better understanding of 
the spin structure of the proton $\cite {ANS}$.  That is,  
\be
{d \over dt} (\alpha_s(t) \Delta g(t))= 0 (\alpha_s(t)^2)
\ee
where $\Delta g(t)$ is the first moment of polarised gluon 
distribution function.  From the above equation 
it is clear that polarised gluonic contribution to
proton spin could be important at very high energies.  But this equation
does not say anything about the absolute value of gluonic contribution.

Now let us consider the first moment of 
the polarised quark fragmentation function into polarised hadron.  
From eqn.(\ref {apeqnq}), it turns out 
\be
{ d \over dt} \Delta D_q^\Lambda(t)={1 \over \pi} 
\alpha_s(t) \Delta D_g^\Lambda(t)
\label{aldq}
\ee
From eqns. (\ref {aldg}) and (\ref {aldq}), we find that 
$\Delta D_{q_i}^\Lambda(t)$ grows as $t$.  
This is to be compared with the first moment of the polarised quarks 
in polarised hadron which is scale independent.  

Hence the relations(eqns.(\ref {aldg},\ref {aldq}))
suggest that at very high
energies the polarised gluon fragmentation into polarised
hadrons(say $\Lambda$) is as significant as polarised quark fragmentation.  
This point is very important when one is interested in
the analysis of integrated asymmetries coming from various partons.  
Also the first moment of the fragmentation
functions will tell us given a parton of definite polarisation,
how much of its polarisation is 
transferred into the produced hadron.  
This is to be compared with the interpretation of the 
first moment of polarised parton 
distributions that it measures how much of hadron's(say proton's) helicity
is shared by the parton.  If we sum over all these contributions,
it will turn out to be the helicity of the hadron. 
Similarly in the fragmentation sector if we sum
over all the hadron fragmentation functions coming from a specific
polarised parton with their appropriate polarisations then 
it will coincide with the polarisation of the parton.

\section{Asymmetries}
The analysis in the last chapter shows that the gluonic contribution to
the polarised $\Lambda$ is as important as the quark contribution.
Hence the QCD corrections to the processes which involve
the extraction of these fragmentation functions are important.
Recently, Burkardt and Jaffe $\cite {BUR}$ have discussed the method of
extracting various polarised quark fragmentations to polarised
$\Lambda$ by measuring some specific asymmetries in both unpolarised
and polarised $e^-$ $e^+$ scattering experiments.  
As a preliminary effort, in the reference $\cite {RAV1}$, $\cite {RAV2}$
various QCD corrections are computed to the cross 
section asymmetry when there is no $Z$ exchange
(purely electromagnetic ($em$)).  The factorisation of IR singularities
has also been verified to order $\alpha_s(Q^2)$ $\cite {RAV1}$.
In the following, we compute the QCD corrections to those asymmetries
(discussed by Burkardt and Jaffe)
which involve both $\gamma$ and $Z$ vector bosons as intermediate particles.

\subsection{\bf The Unpolarised $e^-~e^+~$ Scattering}
First we compute the following asymmetry $\cite {BUR}$.  
\be
\dsc=\sum_a \int_x^1 {dy \over y } \left[\dsd\right] 
\Delta D_a^{\Lambda} (x/y,Q^2)
\label{DSC}
\ee
In the above asymmetry the spins of the initial leptons are averaged.  
This asymmetry is identically zero when there is 
only $\gamma$ exchange.  At very high
energies, the $Z$ exchange is also possible.  This will make
the asymmetry non-zero thanks to parity violation.  In the above
equation $\Omega$ is the solid angle within which the produced
parton fragments into $\Lambda$.  The arrows in the parenthesis of
$\Lambda$ and the partons denote their helicities with respect to
the beam direction.  The scale $Q^2$ is the invariant mass of the photon
or $Z$ produced(i.e.,$q^2=Q^2$) at the $e^+~e^-$ vertex.  The sum is taken over
all the partons such as quarks, anti-quarks and gluons fragmenting into
$\Lambda$.  Recall that the kinematic scaling variables $x$ and $y$ are 
defined as $x=2 p_\Lambda.q/Q^2$ and $y=2 p_a.q/Q^2$ respectively.  
The parton level asymmetry given in the eqn.(\ref {DSC}) is factorisable as
\be
\dsd= {1 \over 4 Q} \sum_{I=Z,\gamma Z} {\cal L}^{(I)}_{\mu \nu}(Q^2)
{\cal D}_{(I)}(Q^2) \sum_{j=1,3,4} {\cal T}_j^{\mu \nu} H^{(I)a}_j(y,Q^2)
\label{GZ}
\ee
where ${\cal D}_{(I)}(Q^2)$ are propagators given by
\be \begin{array}{rl}
{\cal D}_{(Z)}(Q^2)={\displaystyle 1 \over \displaystyle (Q^2 - M_Z^2)^2} \quad & \quad
{\cal D}_{(\gamma Z)}(Q^2)={\displaystyle 1 \over\displaystyle 
Q^2 (Q^2 - M_Z^2)}
\end{array}
\ee
The tensors ${\cal T}_j^{\mu \nu}$ are constructed by looking at the
symmetry properties of the amplitudes for direct $Z$ and $\gamma Z$
interference contributions.  These two amplitudes are individually
gauge invariant in the mass less limit as can be checked explicitly.  Hence
these tensors are parametrised as 
\be \begin{array}{rl}
{\cal T}_1^{\mu \nu} = {\displaystyle i \over\displaystyle p_a.q} 
\epsilon^{\mu \nu \lambda \sigma} q_\lambda
p_{a\sigma}  \quad& \quad
{\cal T}_3^{\mu \nu} = -g^{\mu \nu} + {\displaystyle q^\mu\displaystyle q^\nu
\over \displaystyle Q^2}
\nonumber
\end{array}
\ee
\be
{\cal T}_4^{\mu \nu} ={\displaystyle 1 \over p_a \cdot q }
\left(p_a^\mu - q^\mu {
\displaystyle p_a.q \over \displaystyle  Q^2}\right)
\left(p_a^\nu - q^\nu { \displaystyle p_a.q \over \displaystyle  Q^2}\right) 
\nonumber
\label{tengz}
\ee
The tensors proportional to $q^\mu$ in ${\cal T}_3$ and ${\cal T}_4$ 
are immaterial
as the leptonic tensors are individually gauge invariant.  
The leptonic tensors ${\cal L}_{(I)}^{\mu \nu}$ can be easily workout
and are found to be
\bea
{\cal L}_{(Z)}^{\mu \nu}&=& {  \pi \alpha \over Sin^2\theta_W Cos^2\theta_W}
\left [(v_e^2+a_e^2) l^{\mu \nu} - 2 i v_e a_e \tilde  l^{\mu \nu}
\right] \nonumber \\ 
{\cal L}_{(\gamma Z)}^{\mu \nu}&=& { 4 \pi \alpha \over Sin\theta_W 
Cos\theta_W}
\left [v_e  l^{\mu \nu} -  i a_e \tilde  l^{\mu \nu}
\right] 
\label{lepgz}
\eea
where
\be \begin{array}{rl}
{ l}^{\mu \nu} = q_1^\mu q_2^\nu + q_1^\nu q_2^\mu - g^{\mu \nu} q_1.q_2
\quad &
\tilde { l}^{\mu \nu} = \epsilon^{\mu \nu \alpha \beta} q_{1 \alpha}
q_{2 \beta}
\end{array}
\nonumber
\ee
with $q_1$ and $q_2$ being the momenta of incoming positron and electron
respectively.  Here $\alpha$ is the fine structure constant, 
$v_e$ and $a_e$ are the vector and axial vector couplings
in the  $e^+ e^- Z$ vertex and $\theta_W$ is the Weignberg angle\cite{PDB}.
It is clear from the above tensors that the terms proportional 
to $q^\mu$ or $q^\nu$ in the tensors ${\cal T}_3$ and ${\cal T}_4$ when 
contracted with the leptonic tensors give zero contribution.  

Substituting the tensors (eqn.(\ref {tengz})) and the leptonic
tensors (eqn.(\ref {lepgz})) in the eqn.(\ref {GZ}), we obtain
\bea
\dsd\!\! \!&=&\!\!\!{ 1 \over 2 Q^2} { \pi \alpha \over Sin\theta_W Cos\theta_W}
{1 \over (Q^2 - M_Z^2)} 
\left [ Cos\theta {\cal H}_1^a 
+ {\cal H}_3^a + {y \over 4} (1- Cos^2 \theta)
{\cal H}_4^a \right] 
\eea
where the general form of ${\cal H}_i^a$ for $i=3,4$ is given by
\be
{\cal H}_i^a= {Q^3 \over Q^2 -M_Z^2} {1 \over 2 Sin\theta_W Cos\theta_W}
(v_e^2+a_e^2) H_i^{(Z)a} + 2 Q v_e H_i^{(\gamma Z)a}
\ee 
and that of ${\cal H}_1^a$ is given by
\be
{\cal H}_1^a={Q^3 \over Q^2 - M_Z^2 }{ 1 \over Sin\theta_W Cos\theta_W}
v_e a_e H_1^{(Z)a} + 2 Q a_e H_1^{(\gamma Z)a}
\ee
where, $v_q$ and $a_q$ are vector
and axial vector couplings in $q \bar q Z $ vertex \cite {PDB}.  
The superscript in $H$ denotes
the origins of these contributions such viz, direct $Z$ or the interference
between $\gamma$ and $Z$ contributions.   
The contribution to $H_i^{(I)a}$ can come from processes to lowest order
(see fig.1)($\alpha_s^{(0)}$) in $\alpha_s$ (no gluon emission) as well as
from processes involving single gluon emission(see fig.2) 
and virtual contributions(see fig.3)  
to first order in $\alpha_s$.  The evaluation
of asymmetries to lowest order is very simple as it involves only
two body phase space.  When the gluon accompanies the quark and anti-quark
pair, then the evaluation is cumbersome due to three body
phase space integral.  We have given below a simple looking formula
to compute the three body phase space after performing most of
the integrals using the delta functions.  
The $H_i^{(I)a}~$s are related to the matrix elements as follows:
\be
H_i^{(I)a}={\displaystyle Q \over \displaystyle 32 (2 \pi)^3} \int dx_1 
{\cal P}^{\mu \nu}_i {\displaystyle 1 \over\displaystyle  4 \pi} |M_I^a|^2_{\mu \nu}
(\downarrow -\uparrow) 
\ee
where the projectors are given by
\bea
{\cal P}_1^{\mu \nu}&=& i \epsilon^{\mu \nu \lambda \sigma}{\displaystyle 
p_{a\lambda} q_\sigma  \over \displaystyle 2 p_a.q}\nonumber\\
{\cal P}_3^{\mu \nu}&=&-{1 \over 2} \left( g^{\mu \nu}+ 4{\displaystyle
p_a^\mu p_a^\nu \over Q^2 y^2} \right)\nonumber \\
{\cal P}_4^{\mu \nu}&=&{1 \over y} \left( g^{\mu \nu}+12{\displaystyle
p_a^\mu p_a^\nu \over Q^2 y^2} \right)
\eea
The terms $H_i^{(I)a}$ can be computed from the fig.2 and are found to
be of the form
\be \begin{array}{rl}
H_i^{(Z)a}= 3 {\displaystyle \alpha Q \over\displaystyle  64 \pi Sin^2\theta_W Cos^2\theta_W} C_i^{(Z)a};
&
H_i^{(\gamma Z)a}= -3 {\displaystyle \alpha e_q Q \over \displaystyle 16 \pi Sin\theta_W Cos\theta_W} 
C_i^{(\gamma Z)a}
\label{HI}
\end{array}
\ee
where 
\be \begin{array}{rl}
C_1^{(Z)a} = {1 \over 2} (v_q^2+a_q^2) {\cal C}_1^{a} \quad
& \quad C_1^{(\gamma Z)a} =  v_q ~{\cal C}_1^{a} \\
\quad \quad C_i^{(Z)a} = \eta_a v_q a_q ~{\cal C}_i^{a} \quad \quad  
& \quad C_i^{(\gamma Z)a} = \eta_a a_q ~{\cal C}_i^{a}   \\
\end{array}
\label{CI}
\ee
with $i=3,4$ and $\eta_q=1$ for quarks and $-1$ for anti-quarks.  
The factor $3$ appearing
in the eqn.(\ref {HI}) is due to the number of colours.  The functions
${\cal C}_i^{a}$ are computed in the appendix.  
As we have already mentioned the functions ${\cal C}_i^a$ to lowest
order get contributions from the $Z$ or the $\gamma Z$ decaying
into a polarised quark and antiquark pair.  This asymmetry is just
proportional to $\delta(1-y)$ but for charge and other group
factors.  To order $\alpha_s$, ${\cal C}_i^a~$s get contributions from
two types of processes: 1. polarised quark or antiquark production 
with a real gluon emission and virtual gluon corrections to the quark and
anti-quark(self energy and vertex corrections) 2. polarised 
gluon emission from unpolarised quark and anti-quark pair.  
These processes suffer 
Infrared(IR) divergences when masses of the quarks and gluons
are taken to be zero.  In order to regulate these divergences we give
gluons a small mass $m_g$ and finally take the limit $m_g$ going to zero.
We compute these cross sections in the limit $p_a.q$ and $Q^2$ 
tending to infinity with their ratio fixed.  This is analogous to 
DIS limit but the scaling variable here is inverse of the Bjorken
scaling variable.

\bea
{\cal C}_3^a&=&\delta_{a,q/\bar q} \delta(1-y) + {4 \over 3}
{\alpha_s \over 2 \pi} \left ( C_G^a- {4 \over y^2} C_P^a\right)\nonumber\\
{\cal C}_4^a&=&-2 \delta_{a,q/\bar q} \delta(1-y) + {4 \over 3}
{\alpha_s \over 2 \pi} \left (-{2 \over y}C_G^a+ {24 \over y^3} C_P^a\right)
\nonumber \\
{\cal C}_1^a&=&\delta_{a,q/\bar q} \delta(1-y) + {4 \over 3}
{\alpha_s \over 2 \pi} C_1^a
\label{}
\eea

Notice that the delta function is absent for gluons.
The functions $C_i^a$ for quarks are given below,
\bea
C_1^q&=& \left ( {1+ y^2 \over 1-y} \right)_+ \log 
\left ({Q^2 \over m_g^2}\right) 
+ (1+ y^2) \left ( {\log(1-y) \over 1-y} \right)_+ +
{1+y^2 \over 1-y}\log(y) \nonumber \\
&&
-{3 \over 2} (1-y)
-{3 \over 2} \left( 1 \over 1-y\right)_+ -\left({9\over 4}-
{\pi^2 \over 3} \right) \delta(1-y) \nonumber \\
C_G^q&=&C_1^q + 2-y \nonumber \\
C_P^q&=&{y^2 \over 4}
\label{}
\eea 
Notice that the above expressions are well defined in soft limit.
As is shown in the appendix, the soft singularities coming from real
gluon emission diagrams are exactly cancelled by that
coming from the virtual diagrams such as self energy and vertex corrections.
The term $\log (m_g^2)$ purely comes from the collinear divergence
which can not be avoided in the mass less limit.  This ill defined 
term will finally be absorbed into the fragmentation functions at the
level of $\Lambda$ production cross section.  Hence the fragmentation
functions are defined in what is usually called the massive gluon scheme. 
Similarly for gluons we have
\bea
C_1^g&=& 2 (2-y) \log \left ({Q^2 y^2 \over m_g^2}\right) - 4(2-y)
\nonumber\\
C_G^g&=&0
\nonumber\\
C_P^g&=&0
\label{glco}
\eea
The asymmetry given in the eqn.(\ref {DSC}) is expressed in terms of the
${\cal C}_i^{(I)a}~$s as

\bea
\dsc\!\!\!\!\!&=&\!\!\! 3 { \alpha^2 \over 2 Q^2}\!\sum_{a=q,\bar q, g} \left[ 
-2 \chi_1\!v_e a_q e_q \eta_q  \left({\cal C}_3^{a} \!+\! 
{x \over 4} (1\!-\! Cos^2\theta) {\cal C}_4^{a}
\right)\right. \nonumber \\
&&\left. +2 \chi_2 (v_e^2+a_e^2) v_q a_q \eta_q 
\left({\cal C}_3^{a} + {x \over 4} (1- Cos^2\theta)
{\cal C}_4^{a}\right)\right. \nonumber \\
&&\left. -2 Cos\theta  \left(\chi_1 e_q v_q a_e {\cal C}_1^{a} -
\chi_2 a_e v_e (v_q^2\!+\!a_q^2) {\cal C}_1^{a}\right)\right] \otimes  
\Delta D_a^{\Lambda}(x) \nonumber \\
\label{asymone}
\eea
where 
\be \begin{array}{rl}
\chi_1 ={\displaystyle Q^2 \over \displaystyle Q^2-M_Z^2}
{\displaystyle 1 \over\displaystyle  16 Sin^2\theta_W Cos^2\theta_W};
 & 
\chi_2 = {\displaystyle Q^4 \over\displaystyle  (Q^2-M_Z^2)^2}
{\displaystyle  1 \over \displaystyle 256 Sin^4\theta_W
Cos^4\theta_W} 
\end{array}
\ee
The $\chi_1$ and $\chi_2$ for $Z$ propagator with finite width 
will be modified and are found in ref. \cite {PDB}  
and the convolution of two functions $f(x)\otimes g(x)$ means
\be
f(x) \otimes g(x) = \int_x^1 {dy \over y} f(y) g(x/y) 
\ee
\subsection{\bf The Polarised $e^-~e^+$ Scattering}
Now let us find out the following asymmetry $\cite {BUR}$ where one of the
initial leptons and the produced parton are polarised 
\be
\dsg=\sum_a \int_x^1 {dy \over y } \left[\dsf\right] 
\Delta D_a^{\Lambda} (x/y,Q^2)
\label{DSG}
\ee
Following the similar procedure, we can decompose the parton
level asymmetry in terms of leptonic(${\cal L}^{\mu \nu}$) and 
gauge invariant hadronic(${\cal T}^{\mu \nu}$)tensors as
\be
\dsf={1\over 2 Q} \sum_{I=\gamma,Z,\gamma Z} {\cal L}^{(I)}_{\mu \nu}
(\downarrow)
{\cal D}_{(I)}(Q^2) \sum_{j=1,3,4} {\cal T}_j^{\mu \nu} H^{(I)a}_j(y,Q^2)
\label{GGZ}
\ee
where the new ${\cal D}_{(\gamma)}(Q^2)= 1/Q^4$.  In the above equation,
$I=\gamma$ corresponds to the diagrams where only photons are
intermediate vector bosons.  This is the extra contribution which is
absent in the previous asymmetry.  This is possible because of
the polarisation of initial lepton.  This diagram is gauge invariant as it is.
The diagrams for $I=Z,\gamma Z$ are discussed already.
Notice that the diagrams are split in such a way that for each $I$
the diagrams are individually gauge invariant, hence the simple
tensor decomposition.   
The leptonic tensors ${\cal L}_{(I)}^{\mu \nu}(\downarrow)$ for the polarised 
electron are found to be
\bea
{\cal L}_{(\gamma)}^{\mu \nu}(\downarrow) &=& 8 \pi \alpha 
\left[ l^{\mu \nu} -i \tilde l^{\mu \nu} \right]\nonumber \\
{\cal L}_{(Z)}^{\mu \nu}(\downarrow) &=& {\displaystyle \pi \alpha 
\over \displaystyle 2 Sin^2\theta_W Cos^2\theta_W}(v_e+a_e)^2 
\left[ l^{\mu \nu} -i \tilde l^{\mu \nu} \right]\nonumber \\
{\cal L}_{(\gamma Z)}^{\mu \nu}(\downarrow) &=& {\displaystyle 2 \pi \alpha 
\over \displaystyle  Sin\theta_W Cos\theta_W}(v_e+a_e) 
\left[ l^{\mu \nu} -i \tilde l^{\mu \nu} \right]
\label{lepggz}
\eea
where $l^{\mu \nu}$ and $\tilde l^{\mu \nu}$ are already given.  
Substituting the above
leptonic tensors in eqn.(\ref{GGZ}) we find,
\bea
\dsf\!\!\!\!&=&\!\!\! {\displaystyle 1 \over \displaystyle 2 Q^2}
{\displaystyle \pi \alpha \over \displaystyle Sin\theta_W Cos\theta_W}
{\displaystyle (v_e+ a_e) \over \displaystyle Q^2-M_Z^2} 
\left[{\cal H}_3^a(\downarrow)\!+\!Cos\theta {\cal H}_1^a(\downarrow)
\right.\nonumber \\
&&\!\!\left. +{y \over 4}(1- Cos^2\theta){\cal H}_4^a(\downarrow)\right] 
+ Cos\theta {\displaystyle 4 \pi \alpha \over Q^3} {\cal H}_{\gamma}^a
(\downarrow)
\label{DSE}
\eea
The general form of ${\cal H}_i^a(\downarrow)$ is found to be
\be
{\cal H}_i^a(\downarrow)= {\displaystyle Q^3 \over \displaystyle 
Q^2 -M_Z^2} {\displaystyle 1 \over \displaystyle 2 Sin\theta_W
Cos\theta_W} (v_e+a_e) H_i^{(Z)a} + 2 Q H_i^{(\gamma Z)a}
\label{HDA}
\ee
The functions $H_i^{(I)a}$ for $i=1,3,4$ are given in the
eqns.(\ref {HI}).  The new ${\cal H}_{\gamma}^a(\downarrow)$
is given by
\be
{\cal H}^a_{\gamma}(\downarrow)=3 {\displaystyle \alpha  Q \over
\displaystyle 8 \pi} e_q^2 {\cal C}^{a}_1
\label{NHG}
\ee

Substituting eqn.(\ref {DSE}) in eqn.(\ref {DSG}) and using
the eqns.(\ref {HDA},\ref{NHG},\ref{HI},\ref{CI}) we find
\bea
\dsg\!\!\!\!&=&\!\!\!3 {\alpha^2 \over 2 Q^2}\! \sum_{q,\bar q,g}
\left[\!- \chi_1 (v_e\!\!+\!a_e) 
a_q e_q \eta_q\! \left (2{\cal C}_3^{a}
\!\!+\!\!{x \over 2} (1\!-\! Cos^2\theta){\cal C}_4^{a}
\right) \right. \nonumber \\
&& \left. +~  \chi_2~ (v_e+a_e)^2 v_q a_q \eta_q \left(2{\cal C}_3^{a}
+{x \over 2} (1- Cos^2\theta){\cal C}_4^{a} \right) \right.
\nonumber \\
&&\left.  -2~ \chi_1~ (v_e+a_e) v_q e_q~ Cos\theta~ {\cal C}_1^{a}
\right. \nonumber\\
&&\left. +  \chi_2~ (v_e+a_e)^2 (v_q^2+a_q^2)~ Cos\theta~~ {\cal C}_1^{a}
\right. \nonumber \\
&& \left. +~e_q^2~ Cos\theta ~{\cal C}^{a}_1
\right]\otimes \Delta D_a^\Lambda(x)
\label{asymtwo}
\eea
From the eqns.(\ref {asymone}, \ref {asymtwo}) we find that the 
asymmetries reduce to those given Burkardt and Jaffe(BJ)'s paper
$\cite {BUR}$ when we 
put the strong coupling
constant $\alpha_s(Q^2)$ equal to zero confirming 
the correctness of our results presented in this paper.
Also, we can reproduce the results for the asymmetries in the absence of
$Z$ exchange diagrams by putting
both $\chi_1$ and $\chi_2$ equal to zero.  Hence our results presented
here are consistent with BJ asymmetries.  The extraction
of both quark and gluons now becomes complicated because of the presense
of gluonic fragmentation functions.  Notice that the expressions 
(eqns. (\ref {asymone}), (\ref {asymtwo})) look very similar to
that of BJ asymmetries with the replacement of quark and antiquark fragmentation
functions by the parton fragmentation functions convoluted with 
complicated looking functions ${\cal C}_i^a$. 
This simple fact that the structures are identical 
might help us in the extraction of
the combination of fragmentation functions with ${\cal C}_i^a$
instead of just the fragmentation functions.  From this combination
and an independent measurement on polarised gluonic contribution
one should be able to disentangle both the quark and gluon fragmentation
functions.  Once these functions are measured in the laboratory,
these have nice physical interpretation that their first moments
will tell us the polarisation contribution coming from various partons to
a specific hadron in the sense described earlier.  As is seen 
from the analysis of $Q^2$ evolutions of 
these fragmentation functions using AP evolution equations, 
the first moment of
quark and gluon fragmentation functions will play a crucial role
at high energies in the production of polarised $\Lambda$ hyperons
at $e^-~e^+$ annihilation processes.  
 
\section{Conclusion}
In this paper we have extensively studied the polarised $\Lambda$ 
production in $e^+~e^-$ annihilation process.  This involves the
measurement of what are called polarised fragmentation functions
of quarks and gluons into polarised $\Lambda$. 
The extraction of the fragmentation functions of different
flavours is very complicated due to the charge factors multipling 
them.  Burkardt and Jaffe were successful in disentangling
these distributions by looking at various asymmetries at different
kinematical regions.  In this paper, using the Altarelli-Parisi 
evolution equation, we have shown
that the polarised gluon fragmentation is also important when one
is interested in finding spin or helicity 
coming from various partons to the produced polarised $\Lambda$.
In fact we find that the first moment of the polarised gluon fragmentation
is logarithmically raising as $Q^2$ increases.  So at high
energies gluons play crucial role invalidating the naive expectations
based on simple coupling constant arguments.  This is similar
to the behaviour one encounters in the study of spin content of
the proton or any hadron.  In fact the behaviour in the fragmentation sector
is much more stronger than that in the  structure function sector
due to the vanishing of first moment of $\Delta P_{qg}(x)$. 
Since the gluons can contribute only from order $\alpha_s(Q^2)$,
we have considered the full QCD corrections to the asymmetries
discussed earlier.   This of course, further complicates the
extraction procedure.   The phenomenological implications
of these QCD corrections and the extraction of gluonic fragmentation
function are under investigation.  
It is worth to recollect the status of gluonic contribution to polarised
structure function measured in DIS experiments.  The status
here is very similar to that in DIS due to the scheme dependence of
these gluonic contributions.  Hence, a scheme independent
measurement of polarised gluonic fragmentation function
would substantiate our analysis.  The extraction of these distribution
is very important primarily due to two reasons.  One of them is that
this might tell more about quark and gluon dynamics in the fragmentation
sector.  Particularly, the measurement of non strange quark contributions
to polarised $\Lambda$ production will be a test of existing models for
polarised $\Lambda$ fragmentation.  Also, the importance of polarised
gluons in the $\Lambda$ production can be experimentally tested.  
 
I would like to thank prof. M.V.N. Murthy for his constant encouragement.  
It is a pleasure to thank Prof. R.L. Jaffe for his critical 
comments on the part of the work presented here.
 
\appendix
\section{\bf Matrix Elements and Integrals}
In this appendix we outline the evaluation $H_i^{(I)a}$ in the massive
gluon scheme.  We have split the amplitude into three parts.
That is, 1. $\gamma \rightarrow q \bar q g$, 2. $Z \rightarrow q \bar q g$ and
3. the interference of $\gamma$ and $Z$ initiated processes.
All three are independently gauge invariant.  

Let us first consider $\gamma \rightarrow q \bar q g$ process.
If the quark is polarised then there will be virtual contributions
due to gluons in addition to gluon emission diagrams.
These diagrams are both soft and collinear divergent.
In addition, the virtual diagrams are ultraviolet divergent(UV).
The soft and collinear divergences are regulated by giving a small
mass to the gluon.  The UV divergence is regulated using
Pauli-Villars regularisation.  

To produce polarised quark, unpolarised anti-quark and gluon
the initial virtual photon should be polarised, otherwise the
amplitude is identically zero.  We project out this contribution
as
\be
i \epsilon^{\mu \nu \lambda \sigma} {\displaystyle p_{2\lambda} q_\sigma
\over \displaystyle 2 p_2.q}|M^q_\gamma|^2_{\mu \nu}(\downarrow-\uparrow) 
= 256 \alpha \alpha_s e_q^2 \pi^2 {\cal M}^q_1
\ee
where the momenta of the particles are given in the fig.2, 
the Mandelstam variables are $s=(p_1+p_3)^2, t=(p_2+p_3)^2$ 
and $u=(p_1+p_2)^2$ satisfying $s+t+u=Q^2+m_g^2$.  Here, 
\bea
{\cal M}^q_1&=&(s t - m_g^2 Q^2) \left ( {s+2 t -2 m_g^2 -Q^2 \over
t^2 (s-Q^2) } + {1 \over s^2}\right)\nonumber \\
&& + 2 (s+t -m_g^2-Q^2) {Q^2 s+s t- m_g^2 Q^2-Q^4 \over s t (Q^2-s) }
\label{M1}
\eea

The processes involving only $Z$ particles are given in the fig.2.
Due to the axial coupling in addition to the vector coupling, 
polarised particles can be produced from the virtual $Z$ decay even if one
sums the polarisations of $Z$ unlike the case of
$\gamma$.  Hence we project out the following contributions from
the matrix element when the quark is polarised:
\bea
-g^{\mu \nu} |M^q_Z|^2_{\mu \nu}(\downarrow-\uparrow)&=& {64 \pi^2 
\alpha \alpha_s  \over Cos^2\theta_W
Sin^2\theta_W } v_q a_q {\cal M}^q_G\nonumber \\
{\displaystyle p_2^\mu p_2^\nu \over \displaystyle Q^2}|M^q_Z|^2_{\mu \nu}(\downarrow-\uparrow)&=&
{32 \pi^2 \alpha \alpha_s \over Cos^2\theta_W
Sin^2\theta_W} v_q a_q {\cal M}^q_P\nonumber \\
i \epsilon^{\mu \nu \lambda \sigma} {\displaystyle p_{2\lambda} q_\sigma
\over \displaystyle 2 p_2.q}|M^q_Z|^2_{\mu \nu}(\downarrow-\uparrow) 
&=& {16 \pi^2\alpha \alpha_s  \over Cos^2\theta_W
Sin^2\theta_W}(v_q^2+a_q^2)  {\cal M}^q_1
\label{MZq}
\eea
\bea
{\cal M}^q_G &=&(s t - m_g^2 Q^2) \left ( {1 \over s^2} + {1 \over t^2} \right)
+ 2 (m_g^2+Q^2) { m_g^2+Q^2-s-t \over s t}
\nonumber \\
{\cal M}^q_P&=&(m_g^2-t)^2 { m_g^2 +Q^2 -s-t \over t^2 Q^2}
\label{Mq}
\eea
Here ${\cal M}^q_1$ is already given in the eqn.(\ref{M1}).

Next we compute the interference term coming from $Z$ and $\gamma$ decaying
into polarised quark, anti-quark and gluon 

\bea
-g^{\mu \nu} |M^q_{\gamma Z}|^2_{\mu \nu}(\downarrow-\uparrow)&=&- {256 \pi^2 
\alpha \alpha_s e_q \over Cos\theta_W
Sin\theta_W } a_q {\cal M}^q_G\nonumber \\ 
{\displaystyle p_2^\mu p_2^\nu \over \displaystyle Q^2} |M^q_{\gamma Z}|^2_{\mu \nu}(\downarrow-\uparrow)&=&
-{128 \pi^2 \alpha \alpha_s e_q\over Cos\theta_W
Sin\theta_W} a_q {\cal M}^q_P\nonumber \\
i \epsilon^{\mu \nu \lambda \sigma} {\displaystyle p_{2\lambda} q_\sigma
\over \displaystyle 2 p_2.q}|M^q_{\gamma Z}|^2_{\mu \nu}(\downarrow-\uparrow) 
&=& -{128 \pi^2 \alpha \alpha_s e_q \over Cos\theta_W Sin\theta_W}v_q 
{\cal M}^q_1
\eea
Noting that $s=Q^2(1-x_2)$ and $t=Q^2(1-x_1)$ where $x_i=2 p_i.q/Q^2$
the relevant integrals are presented below
\bea
\int dx_1 {\cal M}_1 &=&\left({ 1 +x_2 \over 1 -x_2 }\right) \left (
\log(x_2 (1-x_2) + \log\left({Q^2 \over m_g^2}\right)\right)
\nonumber \\
&&-{3 \over 2}
(1-x_2)- {3 \over2} {1 \over 1 - x_2} +{5 \over 4} \delta(1-x_2)
\nonumber\\
\int dx_1 {\cal M}_G &=& \int dx_1 {\cal M}_1 + 2 - x_2
\nonumber \\
\int dx_1 {\cal M}_P&=&{x_2^2 \over 2}
\label{intt}
\eea
The virtual contributions can be computed from the diagrams given in the
fig.3.  Here we state the expressions for the amplitudes involving
self energy correction and vertex correction (for general vertex 
of the type $(a+ b \gamma_5)\gamma_\mu$).
Both diagrams(self energy and vertex correction) in fig.3 
are UV divergent separately.
As we have already mentioned, we regulate these UV divergences 
in Pauli-Villars regularisation scheme.  They also suffer 
from soft divergences when we set all the masses to
be zero.  Here also, we give a small mass to the gluon to regulate
them.  The amplitude for the self energy insertion
after properly taking care of wave function renormalisation
turns out to be(self energies for both quark and antiquark)
\be
M_\mu^{self}= {\alpha_s \over  \pi} \log\left({m_g^2 \over L}\right)
\bar v_{s'}(p_1)(a+b \gamma_5) \gamma_\mu u_{s}(p_2)
\label{self}
\ee
where  $L$ is the Pauli-Villars regulator to regulate the UV divergence.
$s$ and $s'$ are the polarisations of quark and antiquark respectively.

Similarly the amplitude for the vertex correction turns out to be
\bea
M_\mu^{vertex}&=&-{\alpha_s \over  \pi}\left[\log^2\left({m_g^2 \over Q^2}\right)
+3 \log\left({m_g^2 \over Q^2} \right)
-\log\left({L \over m_g^2}\right)
\right. \nonumber \\
&& \left.+ {7 \over 2} -{\pi^2  \over 3}
\right] \bar v_{s'}(p_1) 
(a+ b \gamma_5) \gamma_\mu u_{s}(p_2)
\label{vert}
\eea
Though the above amplitudes suffer from UV divergence,
when we sum these two amplitudes, we find that
the UV regulator cancels between self energy and vertex corrections, 
thanks to Ward identity. 

Now let us find out the projected matrix elements similar to 
eqns.(\ref {M1}), (\ref {Mq}) when the gluon is polarised
and quark and antiquarks are unpolarised.  We here list ${\cal M}_i^g$
analogous to ${\cal M}_i^q$ (with $p_2$ replaced by $p_3$) 
\bea
{\cal M}_1^g&=&{\eta\over 2}{s t - m_g^2 Q^2 \over (s+t)^2} \left[ 
2 {(s+t) (s+t-m_g^2-Q^2) \over s t} 
+ {4 m_g^2 Q^2 - 2 m_g^2 s - 2 Q^2 s + s^2 -t^2 \over t^2} 
\right.
\nonumber \\
&&\left. + {4 m_g^2 Q^2 - 2 m_g^2 t - 2 Q^2 t + t^2 -s^2 \over s^2}
\right]\nonumber\\
{\cal M}_G^g&=&{\eta\over 2} {1 \over s+t } \left [2 m_g^2(s+t-m_g^2 -Q^2) {t-s \over st}
+(2 Q^2 -s -t) (s t -m_g^2 Q^2) \left( {1 \over s^2} - { 1 \over t^2}
\right) \right]\nonumber \\
{\cal M}_P^g&=&{\eta\over 2 Q^2} {m_g^2 (m_g^2 +Q^2 -s -t) \over s+t } \left [ 2 m_g^2
 {t-s \over st} - {2 m_g^2 Q^2 -m_g^2 s+m_g^2 t - 2 s t \over
t^2}\right. \nonumber \\
&&\left. +{2 m_g^2 Q^2 +m_g^2 s -m_g^2 t - 2 s t  \over s^2}  \right] 
\label{}
\eea
Where $\eta=1$ for $\gamma$ or $Z$ diagrams
and $\eta=2$ for $\gamma Z$ interference terms.
We find that the results given in eqn. (\ref {glco}) can be got by 
integrating the above matrix elements with respect to $x_2$. 
The integrations of the matrix elements ${\cal M}_G^g$ and ${\cal M}_P^g$ 
are identically zero in the limit $m_g$ taken to be zero.

\vspace{2cm}
{\bf \Large Figure Captions:}
\begin{enumerate}
\item
Graph contributing to $e^+(q_1) e^- (q_2)\rightarrow \bar q(p_1)
q(p_2)$.
\item
Graphs contributing to $e^+(q_1) e^- (q_2)\rightarrow \bar q(p_1) 
q(p_2) g(p_3)$.
\item
Self energy and vertex corrections to $e^+(q_1) e^- (q_2)
\rightarrow \bar q(p_1) q (p_2)$ 
\end{enumerate}
\end{document}